\documentclass[prb,twocolumn,showpacs]{revtex4}
\usepackage{color,soul}
\usepackage{amsmath}
\usepackage{dcolumn}
\usepackage{graphicx}
\usepackage{bm}
\usepackage{amssymb}
\usepackage{subfigure}
\begin{document}

\title{An exact formula for the optical conductivity of the two dimensional Hubbard model and its application to the cuprate superconductors}
\author{Xinyue Liu and Tao Li}
\affiliation{Department of Physics, Renmin University of China, Beijing 100872, P.R.China}

\begin{abstract}
Understanding the origin of electron incoherence is believed to be the first step toward the resolution of the mysteries of the high-T$_{c}$ cuprate superconductors. Such electron incoherence manifests itself most evidently in the non-Drude form of the optical absorption spectrum of the system. The spectral weight transfer related to such dissipative response, which is absent in conventional Fermi liquid metal, has direct consequence on the dc transport property of the system. However, a theoretical study of the optical conductivity of a strongly correlated model is a formidable task. Here we present an exact formula for the optical conductivity of the 2D Hubbard model from the low energy effective theory perspective. We show that the optical conductivity in Matsubara frequency of the 2D Hubbard model can be represented as the ensemble average of the optical conductivity of non-interacting systems in the background of fluctuating local moment. We find that such an ensemble average can be done exactly with a sign-problem-free Monte Carlo simulation if we assume the widely adopted Millis-Monien-Pines spin susceptibility for the fluctuating local moment. For thermal fluctuation of the local moment, our formula can be used to calculate directly the optical conductivity in real frequency which can be compared with the result of optical measurements in the cuprate superconductors.       
\end{abstract}

\maketitle

\section{Introduction}
An understanding of the non-Fermi liquid behavior is believed to be the first step toward the resolution of the mystery in the high-$T_{c}$ cuprate superconductors. A well known example of such non-Fermi liquid behavior is the non-Drude form of its optical absorption spectrum\cite{Basov,Marel,Heumen}. More specifically, the optical absorption in the cuprate superconductors exhibits a much slower decay with frequency than that in conventional fermi liquid metal and remains substantial even at the energy scale of the band width. It is generally believed that such an anomalous behavior should be attributed to the strong correlation effect in the cuprate superconductors, since neither the electron-phonon coupling nor the impurity scattering is expected to induce optical absorption at such a high energy scale. We note that according to the optical sum rule, a study of electron incoherence at the energy scale of the band width can shed important light on the transport behavior of the system at low energy, for example, the dc resistivity in the normal state and the superfluid density in the superconducting state.

For a purely electronic model, strong electron incoherence is usually attributed to the scattering of the electron with some kind of collective fluctuation of the electron. However, such collective fluctuation usually gain appreciable spectral weight only when the system is close to the transition toward some symmetry breaking phase and thus usually has an energy scale much smaller than the band width. Rather than such soft mode, a broad spectral continuum is needed to understand the origin of the electron incoherence in the cuprates\cite{Conte}. Such a continuum should also be ubiquitous in the phase diagram of the cuprate superconductors, since the non-Drude behavior is not limited to any particular doping.

The fluctuation of the local moment in a doped Mott insulator may just provide such a broad and ubiquitous continuum. In a doped Mott insulator such as the cuprate superconductor, the local moment remains well-defined even when the magnetic long range order is totally suppressed by doping since its existence is protected by the strong electron correlation effect in the system. Indeed, RIXS measurements in the last decade find that spin-wave-like paramagnon fluctuation exists ubiquitously in the phase diagram of the cuprate superconductors\cite{Tacon,Dean}. The energy scale, dispersion, and its integrated spectral weight are found to be almost doping independent. It is very likely that the electron incoherence as manifested in the non-Drude optical absorption behavior has its origin in such ubiquitous paramagnetic fluctuation.

The dual nature of electron in the cuprate as both itinerant quasiparticles and local moments poses a serious challenge to theory. However, at a phenomenological level, one can treat these two kinds of movements as independent degree of freedoms and assume a phenomenological coupling between them. The result is the so called spin-fermion model\cite{Millis,Chubukov}. At a more microscopic level, one can justify such a separation of the electron degree of freedom in the renormalization group perspective. The spin-fermion model has been extensively used in the study of the high T$_{c}$ cuprates. In particular, the theory provides a natural understanding on the origin of the d-wave pairing in the superconducting state.  The scattering from the antiferromagnetic spin fluctuation is also believed to be responsible for the large electron scattering rate and the pseudogap phenomena in the normal state. However, an exact treatment of the effect of the antiferromagnetic spin fluctuation in the spin-fermion model is difficult. For example, the calculation of the optical conductivity of the spin-fermion model is usually done at low perturbative order\cite{Chubukov,Lin}, with the vertex correction treated at different level of accuracy. 

Here we present an exact formula for the optical conductivity of the 2D Hubbard model based on an effective theory description of the fluctuation of the local moment in the system. Our formalism follows closely that adopted in Ref.[\onlinecite{Metzner}]. We find that the optical conductivity of the interacting system can be represented as the ensemble average of the optical conductivity of non-interacting systems in the background of fluctuating local moment. We find further that the simulation of such local moment fluctuation is free from the notorious negative sign problem in either the high temperature limit or the Gaussian limit. In particular, we find that the simulation of the local moment fluctuation in the spin-fermion model is free from the negative sign problem and can be done in a numerical exact manner. We have applied our formula to study the effect of thermal spin fluctuation on the optical conductivity of the cuprate superconductors.  

The paper is organized as follows. In the next section we present a derivation of an exact formula for the optical conductivity of the Hubbard model in terms of an effective theory description of the local moment fluctuation. In the third section, we present an analysis of the negative sign problem in the Monte Carlo simulation of the local moment fluctuation. The fourth section is devoted to a case study on the effect of thermal spin fluctuation on the optical conductivity of the cuprate superconductors. The last section concludes the paper and discuss the possible generalization of the current computation scheme to the study of other physical quantities.  

\section{An exact formula for the optical conductivity of the Hubbard model}
 The Hamiltonian of the Hubbard model studied in this paper is given by
\begin{equation}
H=-\sum_{i,j,\alpha}t_{i,j}c^{\dagger}_{i,\alpha}c_{j,\alpha}+U\sum_{i}n_{i,\uparrow}n_{i,\downarrow}-\mu\sum_{i,\alpha}n_{i,\alpha}
\end{equation}
in which $\alpha=\uparrow,\downarrow$ denotes the spin index of the electron. $t_{i,j}$ denotes the hopping integral between site $i$ and site $j$. $n_{i,\alpha}=c^{\dagger}_{i,\alpha}c_{i,\alpha}$. 

To compute the optical conductivity of the model, we couple the electron to an electromagnetic field through the following Peierls substitution 
\begin{equation}
t_{i,j}\rightarrow \tilde{t}_{i,j}[\mathbf{A}]=t_{i,j}e^{i\mathbf{A}(t)\cdot(\mathbf{r}_{i}-\mathbf{r}_{j})}
\end{equation}
Here $\mathbf{A}(t)$ is the vector potential defined on the bond connecting site $i$  and site $j$. Since we are considering the optical conductivity of the system we can assume that the vector potential is spatially uniform. We have adopted the convention $\hbar=e=c=k_{B}=a=1$ for convenience. Here $a$ is the lattice constant.

Since we will focus on the imaginary time formalism in the following, we will consider 
$\mathbf{A}(\tau)$ instead of $\mathbf{A}(t)$. $\mathbf{A}(\tau)$ can be decomposed in frequency space as
\begin{equation}
\mathbf{A}(\tau)=\sum_{i\omega_{n}}\mathbf{A}(i\omega_{n})e^{-i\omega_{n}\tau}
\end{equation}
in which $\omega_{n}=2n\pi k_{B}T$ denotes the bosonic Mastubara frequency.
The electric current is given by the derivative of the free energy functional $F[\mathbf{A}]=-T\ln Z[\mathbf{A}]$ with respect to the electromagnetic potential
\begin{equation}
\mathbf{J}(i\omega_{n})=-\frac{\partial F[\mathbf{A}]}{\partial \mathbf{A}(-i\omega_{n})}
\end{equation} 
Within the functional path integral formalism, the partition functional $Z[\mathbf{A}]$ can be represented as
\begin{equation}
Z[\mathbf{A}]=\mathrm{Tr} e^{-\beta H[\mathbf{A}]}=\int D[\psi,\psi^{\dagger}] e^{-\mathbf{S}[\psi,\psi^{\dagger},\mathbf{A}]}
\end{equation}
in which $\psi$ and $\psi^{\dagger}$ represent the Grassmannian variables corresponding to the fermion operator $c$ and $c^{\dagger}$, $\beta=1/T$. The action is given by
\begin{equation}
\mathbf{S}[\psi,\psi^{\dagger},\mathbf{A}]=\int_{0}^{\beta} d\tau (\psi^{\dagger}\partial_{\tau}\psi +\mathcal{H}[\psi,\psi^{\dagger},\mathbf{A}])
\end{equation}
Here we have made the abbreviation
\begin{equation} 
\psi^{\dagger}\partial_{\tau}\psi=\sum_{i,\alpha}\psi^{\dagger}_{i,\alpha}\partial_{\tau}\psi_{i,\alpha}
\end{equation} 
The functional $\mathcal{H}[\psi,\psi^{\dagger},\mathbf{A}]$ is given by
\begin{eqnarray}
\mathcal{H}[\psi,\psi^{\dagger},\mathbf{A}]&=&-\sum_{i,j,\alpha}\tilde{t}_{i,j}[\mathbf{A}]\psi^{\dagger}_{i,\alpha}\psi_{j,\alpha}\nonumber\\
&-&\frac{2U}{3}\sum_{i}\mathbf{s}_{i}\cdot\mathbf{s}_{i}-\mu\sum_{i,\alpha}\psi^{\dagger}_{i,\alpha}\psi_{i,\alpha}
\end{eqnarray}
with
\begin{equation}
\mathbf{s}_{i}=\frac{1}{2}\sum_{\alpha,\alpha'}\psi^{\dagger}_{i,\alpha}\bm{\sigma}_{\alpha,\alpha'}\psi_{i,\alpha'},
\end{equation}
Here $\bm{\sigma}$ is the usual spin Pauli matrix. In deriving this equation we have used the following operator identity
\begin{equation}
Un_{i,\uparrow}n_{i,\downarrow}=-\frac{2U}{3}\hat{\mathbf{s}}_{i}\cdot\hat{\mathbf{s}}_{i}+\frac{U}{2}(n_{i,\uparrow}+n_{i,\downarrow})
\end{equation}
where
\begin{equation}
\hat{\mathbf{s}}_{i}=\frac{1}{2}\sum_{\alpha,\alpha'}c^{\dagger}_{i,\alpha}\bm{\sigma}_{\alpha,\alpha'}c_{i,\alpha'}
\end{equation} 
is the spin density operator on site $i$. We have also absorbed the shift of the chemical potential into a redefinition of $\mu$.

The quartic term  in the action can be treated by the Hubbard-Strotonovich(HS) transformation of the following form
\begin{equation}
e^{\frac{2U}{3}\mathbf{s}_{i}\cdot\mathbf{s}_{i}}=\Lambda \int d\vec{\phi}_{i} \ \exp\{-\frac{U}{6}\vec{\phi}_{i}\cdot \vec{\phi}_{i}+\frac{2U}{3}\vec{\phi}_{i}\cdot\mathbf{s}_{i}\}
\end{equation}
in which $\Lambda$ is an unimportant constant. Inserting this expression into Eq.5, we have
\begin{equation}
Z[\mathbf{A}]=\int D[\psi,\psi^{\dagger}] D[\vec{\phi}] e^{-\int_{0}^{\beta} d\tau (\psi^{\dagger}\partial_{\tau}\psi +\mathcal{H}_{\vec{\phi}}[\psi,\psi^{\dagger},\mathbf{A}])}
\end{equation}
in which
\begin{eqnarray}
\mathcal{H}_{\vec{\phi}}[\psi,\psi^{\dagger},\mathbf{A}]&=&-\sum_{i,j,\alpha}\tilde{t}_{i,j}[\mathbf{A}]\psi^{\dagger}_{i,\alpha}\psi_{j,\alpha}\nonumber\\
&-&\frac{2U}{3}\sum_{i}\vec{\phi}_{i}\cdot\mathbf{s}_{i}-\mu\sum_{i,\alpha}\psi^{\dagger}_{i,\alpha}\psi_{i,\alpha}\nonumber\\
&+&\frac{U}{6}\sum_{i}\vec{\phi}_{i}\cdot\vec{\phi}_{i}
\end{eqnarray}

The action is now quadratic in the fermion field $\psi$, which can be integrated out to generate the effective action of the HS field $\vec{\phi}$. It takes the form of
\begin{equation}
Z[\mathbf{A}]=\int D[\vec{\phi}] e^{-\tilde{S}_{eff}[\vec{\phi},\mathbf{A}]}
\end{equation}
with the action given by
\begin{eqnarray}
\tilde{S}_{eff}[\vec{\phi},\mathbf{A}]&=&\int_{0}^{\beta} d\tau  \frac{U}{6}\sum_{i}\vec{\phi}_{i}\cdot\vec{\phi}_{i}\nonumber\\
&-&\mathrm{Tr}\ln([\partial_{\tau}-\mu-\frac{U}{3}\vec{\phi}_{i} \cdot \bm{\sigma}]\delta_{i,j}-\tilde{t}_{i,j}[\mathbf{A}])\ \nonumber\\
\end{eqnarray}
Here the argument of the logarithmic function should be understood as a matrix in the space made up of the direct product of the lattice site, the spin, and the imaginary time. We have omitted possible identity matrix in the space of the imaginary time and spin. 

To derive the electromagnetic response of the system at the linear order, we follow Ref.[\onlinecite{Metzner}] and separate $\tilde{t}_{i,j}[\mathbf{A}]$ into two parts as follows
\begin{equation}
\tilde{t}_{i,j}[\mathbf{A}]=t_{i,j}+v_{i,j}[\mathbf{A}]
\end{equation}
in which 
\begin{equation}
v_{i,j}[\mathbf{A}]=t_{i,j}(e^{i\mathbf{A}(\tau)\cdot(\mathbf{r}_{i}-\mathbf{r}_{j})}-1)
\end{equation}
It can be expanded in $\mathbf{A}$ as follows
\begin{equation}
v_{i,j}[\mathbf{A}]=\sum_{n=1}^{\infty}v^{(n)}_{i,j}
\end{equation}
in which $v^{(n)}_{i,j}$ is the n-th order term in the expansion. For example
\begin{eqnarray}
v^{(1)}_{i,j}&=&it_{i,j}\mathbf{A}\cdot (\mathbf{r}_{i}-\mathbf{r}_{j})=\sum_{i\omega_{n}}\mathbf{j}_{i,j}\cdot\mathbf{A}(i\omega_{n})e^{-i\omega_{n}\tau}\nonumber\\
v^{(2)}_{i,j}&=&-\frac{t_{i,j}}{2}(\mathbf{A}\cdot(\mathbf{r}_{i}-\mathbf{r}_{j}))^{2}\nonumber\\
&=&-\frac{1}{2}\sum_{i\omega_{n},i\omega'_{n}}\mathbf{A}(i\omega_{n})\ \mathbf{K}_{i,j}\mathbf{A}(i\omega'_{n})e^{-i(\omega_{n}+\omega'_{n})\tau} \nonumber\\
\end{eqnarray}
in which 
\begin{eqnarray}
\mathbf{j}_{i,j}&=&it_{i,j}(\mathbf{r}_{i}-\mathbf{r}_{j})\nonumber\\
\mathbf{K}_{i,j}&=&t_{i,j}(\mathbf{r}_{i}-\mathbf{r}_{j})(\mathbf{r}_{i}-\mathbf{r}_{j})
\end{eqnarray}
are the current vector and the inverse effective mass tensor. For later convenience, we define
\begin{eqnarray}
\mathbf{j}_{i,j}(i\omega_{n})=\mathbf{j}_{i,j}e^{-i\omega_{n}\tau}\nonumber\\
\mathbf{K}_{i,j}(i\omega_{n})=\mathbf{K}_{i,j}e^{-i\omega_{n}\tau}
\end{eqnarray}

The effective action $\tilde{S}_{eff}$ can then be rewritten as
\begin{eqnarray}
\tilde{S}_{eff}[\vec{\phi},\mathbf{A}]&=&\int_{0}^{\beta} d\tau  \frac{U}{6}\sum_{i}\vec{\phi}_{i}\cdot\vec{\phi}_{i}\nonumber\\
&-&\mathrm{Tr}\ln(-G^{-1}[\vec{\phi}]-v[\mathbf{A}]) 
\end{eqnarray}
in which
\begin{equation}
G^{-1}_{i\alpha,j\alpha'}[\vec{\phi}]=-(\partial_{\tau}-\mu-\frac{U}{3}\vec{\phi}_{i}\cdot\bm{\sigma}_{\alpha\alpha'})\delta_{i,j}+t_{i,j}
\end{equation}
denotes the inverse Green's function of the electron in the presence of the spin fluctuation field $\vec{\phi}$. We can thus define
\begin{equation}
\tilde{S}_{eff}[\vec{\phi},\mathbf{A}]=S_{eff}[\vec{\phi}]- \mathrm{Tr}\ln(I+G[\vec{\phi}]v[\mathbf{A}])
\end{equation}
in which 
\begin{equation}
S_{eff}[\vec{\phi}]=\int_{0}^{\beta} d\tau   \frac{U}{6}\sum_{i}\vec{\phi}_{i}\cdot\vec{\phi}_{i}-\mathrm{Tr}\ln(-G^{-1}[\vec{\phi}])  
\end{equation}
denotes the action of the spin fluctuation in the absence of the external electromagnetic field. We thus have
\begin{equation}
Z[\mathbf{A}]=\int D[\vec{\phi}] e^{-S_{eff}[\vec{\phi}]}e^{  \mathrm{Tr}\ln(I+G[\vec{\phi}]v[\mathbf{A}]) }
\end{equation}

To calculate the current at the linear order in the electromagnetic field we only need to expand the free energy $F$ to the second order in $\mathbf{A}$. Noting that $v[\mathbf{A}]$ is at least of the first order in $\mathbf{A}$, up to the second order in $\mathbf{A}$ we have
\begin{equation}
\mathrm{Tr}\ln(I+Gv)  \approx \mathrm{Tr}(Gv)-\frac{1}{2}\mathrm{Tr}(Gv)^{2}
\end{equation}
and 
\begin{equation}
e^{\mathrm{Tr}\ln(I+Gv)}\approx 1+\mathrm{Tr}(Gv)-\frac{1}{2}\mathrm{Tr}(Gv)^{2}+\frac{1}{2}(\mathrm{Tr}(Gv))^{2}
\end{equation}
Thus up to the second order in $\mathbf{A}$ the free energy of the system is given by
\begin{eqnarray}
F&=&-T\ln Z\nonumber\\
&\approx&-T\ln \int D[\vec{\phi}] e^{-S_{eff}[\vec{\phi}]} [ 1+\mathrm{Tr}(Gv)\nonumber\\
&-&\frac{1}{2}\mathrm{Tr}(Gv)^{2}+\frac{1}{2}(\mathrm{Tr}(Gv))^{2}]\nonumber\\
&\approx&-T\ln \int D[\vec{\phi}] e^{-S_{eff}[\vec{\phi}]}  [1+\mathrm{Tr}(Gv^{(2)})\nonumber\\
&-&\frac{1}{2}\mathrm{Tr}(Gv^{(1)})^{2}+\frac{1}{2}(\mathrm{Tr}(Gv^{(1)}))^{2}]\nonumber\\
\end{eqnarray}
 Here we have used the result
\begin{equation}
 \int D[\vec{\phi}] e^{-S_{eff}[\vec{\phi}]} \mathrm{Tr}(Gv^{(1)})=0
\end{equation}
This identity can be proved by noting the inversion symmetry of effective action $S_{eff}[\vec{\phi}]$ with respect to $\vec{\phi}$.

Expanding the logarithmic function we arrive at the following formula
\begin{eqnarray}
F&=&-T\ln Z\nonumber\\
&\approx&F[0]-T\ln [ 1+\langle \mathrm{Tr}(Gv^{(2)})\rangle\nonumber\\
&-&\frac{1}{2}\langle \mathrm{Tr}(Gv^{(1)})^{2}\rangle+\frac{1}{2}\langle (\mathrm{Tr}(Gv^{(1)}))^{2}\rangle   ]\nonumber\\
&\approx&F[0]-T\langle \mathrm{Tr}(Gv^{(2)})\rangle\nonumber\\
&+&\frac{1}{2}T\langle \mathrm{Tr}(Gv^{(1)})^{2}\rangle-\frac{1}{2}T\langle (\mathrm{Tr}(Gv^{(1)}))^{2}\rangle
\end{eqnarray}
Here
\begin{equation}
F[0]=-T\ln \int D[\vec{\phi}] e^{-S_{eff}[\vec{\phi}]} 
\end{equation}
is the free energy in the absence of the external field. $\langle O \rangle$ denotes the ensemble average over the fluctuating filed configuration $\vec{\phi}$. It is defined as
\begin{equation}
\langle O \rangle=\frac{\int D[\vec{\phi}] e^{-S_{eff}[\vec{\phi}]} O }{\int D[\vec{\phi}] e^{-S_{eff}[\vec{\phi}]}}
\end{equation} 

The three terms in the last line of Eq.32 are calculated as follows. Inserting the expression of $v^{(2)}$ we have 
\begin{equation}
\langle \mathrm{Tr}(Gv^{(2)})\rangle=-\frac{1}{2}\sum_{i\omega_{n},i\omega'_{n}}\mathbf{A}(i\omega_{n})\langle \mathrm{Tr}[G\mathbf{K}(i\omega_{n}+i\omega_{n'}]\rangle \mathbf{A}(i\omega'_{n})
\end{equation}
Since the system is time translational invariant, we expect
\begin{equation}
\langle \mathrm{Tr}[G\mathbf{K}(i\omega_{n}+i\omega_{n'})]\rangle\propto \delta_{\omega_{n}+\omega'_{n},0} 
\end{equation}
Thus we have
\begin{equation}
\langle \mathrm{Tr}(Gv^{(2)})\rangle=-\frac{1}{2}\sum_{i\omega_{n}}\mathbf{A}(i\omega_{n})\langle \mathrm{Tr}[G\mathbf{K}]\rangle \mathbf{A}(-i\omega_{n})
\end{equation}
Similarly, by inserting the expression of $v^{(1)}$ we have
\begin{equation}
\langle \mathrm{Tr}(Gv^{(1)})^{2}\rangle=\sum_{i\omega_{n},i\omega'_{n}}\mathbf{A}(i\omega_{n})\langle \mathrm{Tr}[G\mathbf{j}(i\omega_{n})G\mathbf{j}(i\omega_{n'})]\rangle \mathbf{A}(i\omega'_{n})
\end{equation}
Using again the time translational symmetry of the system, the expectation value is nonzero only when $\omega+\omega'=0$. We thus have
\begin{equation}
\langle \mathrm{Tr}(Gv^{(1)})^{2}\rangle=\sum_{i\omega_{n}}\mathbf{A}(i\omega_{n})\langle \mathrm{Tr}[G\mathbf{j}(i\omega_{n})G\mathbf{j}(-i\omega_{n})] \rangle \mathbf{A}(-i\omega_{n})
\end{equation}
The last term is given by
\begin{eqnarray}
&&\langle (\mathrm{Tr}(Gv^{(1)}))^{2} \rangle\nonumber\\
&=&\sum_{i\omega_{n},i\omega'_{n}}\mathbf{A}(i\omega_{n})\langle \mathrm{Tr}[G\mathbf{j}(i\omega_{n})]\mathrm{Tr}[G\mathbf{j}(i\omega_{n'})]\rangle \mathbf{A}(i\omega'_{n})\nonumber\\
\end{eqnarray}
This expression is also nonzero only when $\omega_{n}+\omega'_{n}=0$. We thus have
\begin{eqnarray}
\langle (\mathrm{Tr}(Gv^{(1)}))^{2} \rangle&=&\sum_{i\omega_{n}}\mathbf{A}(i\omega_{n})\langle \  \mathrm{Tr}[G\mathbf{j}(i\omega_{n})]\nonumber\\
&\times&\mathrm{Tr}[G\mathbf{j}(-i\omega_{n})]\ \rangle \mathbf{A}(-i\omega_{n})
\end{eqnarray}
Collecting all these three terms we get the expansion of the free energy to the second order in the vector potential as follows
\begin{equation}
F\approx F[0]+\frac{1}{2}\sum_{i\omega_{n}} \mathbf{A}(i\omega_{n})(K+C(i\omega_{n})) \mathbf{A}(-i\omega_{n})
\end{equation}
in which 
\begin{eqnarray}
K&=&T\langle \mathrm{Tr}[G\mathbf{K}]\rangle\nonumber\\
C(i\omega_{n})&=&T\langle \mathrm{Tr}_{c}[G\mathbf{j}(i\omega_{n})G\mathbf{j}(-i\omega_{n})]\rangle
\end{eqnarray}
Here
\begin{eqnarray}
 \mathrm{Tr}_{c}[G\mathbf{j}(i\omega_{n})G\mathbf{j}(-i\omega_{n})]&=&\mathrm{Tr}[G\mathbf{j}(i\omega_{n})G\mathbf{j}(-i\omega_{n})]\nonumber\\
 &-&\mathrm{Tr}[G\mathbf{j}(i\omega_{n})]\mathrm{Tr}[G\mathbf{j}(-i\omega_{n})]\nonumber\\
\end{eqnarray}
denotes the connected part of the current-current correlation function for a given configuration of the fluctuation field $\vec{\phi}$. Here both $K$ and $C(i\omega_{n})$ should be understood as rank-2 tensors.

Differentiating the free energy with respect to the vector potential we get
\begin{equation}
\mathbf{J}(i\omega_{n})=-\frac{\partial F}{\partial \mathbf{A}(-i\omega_{n})}=-(K+C(i\omega_{n})) \mathbf{A}(i\omega_{n})
\end{equation}
The (imaginary frequency)optical conductivity of the system is then given by
\begin{equation}
\sigma(i\omega_{n})=\frac{K+C(i\omega_{n})}{\omega_{n}}
\end{equation} 
Thus the (imaginary frequency)optical conductivity of the interacting system is simply the ensemble average of the optical conductivity of the non-interacting system in the background of fluctuating field $\vec{\phi}$.

\section{Monte Carlo simulation of the optical conductivity of the Hubbard model and the negative sign problem}
Both $K$ and $C(i\omega)$ in Eq.46 can be evaluated in principle if we know the form of $S_{eff}[\vec{\phi}]$, namely, the effective action of the fluctuating moment. For example, this can be done numerically by Monte Carlo sampling over the distribution $e^{-S_{eff}[\vec{\phi}]}$ if it is positive definite. This is generally not true as a result of the notorious negative sign problem in quantum Monte Carlo simulation. However, it can be shown that $e^{-S_{eff}[\vec{\phi}]}$ is positive definite in some important limiting cases that interest us. Here we will assume that the system has a general incommensurate filling so that there is no symmetry to guarantee the exact cancelation of the negative sign problem\cite{Hirsch}.

The first interesting case is the high temperature limit in which $\beta\rightarrow 0$. We can now ignore the time dependence of the fluctuation field $\vec{\phi}$. In such a case, $S_{eff}[\vec{\phi}]$ is nothing but the free energy of the band electron in the background of a static Zeeman field $\frac{2U}{3}\vec{\phi}_{i}$, which is obviously a real number. Thus an accurate estimate of the optical conductivity of the Hubbard model can be obtained at sufficiently high temperature.

The second interesting case is the Gaussian limit when we retain only the quadratic term in the expansion of the effective action $S_{eff}[\vec{\phi}]$ in $\vec{\phi}_{i}$. More specifically, since
\begin{eqnarray}
&&\mathrm{Tr}\ln(-G^{-1}[\vec{\phi}]) =\mathrm{Tr}\ln(-G_{0}^{-1}-\frac{2U}{3}\vec{\phi}_{i}\cdot \mathbf{s}_{i}) \nonumber\\
&=&\mathrm{Tr}\ln(-G_{0}^{-1})+\mathrm{Tr}\ln[\mathbf{I}+\frac{2U}{3}G_{0}\vec{\phi}_{i}\cdot \mathbf{s}_{i}] \nonumber\\
&=&\mathrm{Tr}\ln(-G_{0}^{-1})-\sum_{n}\frac{(-2U/3)^{n}}{n}\mathrm{Tr}[G_{0}\vec{\phi}_{i}\cdot \mathbf{s}_{i}]^{n}\nonumber\\
\end{eqnarray}
to second order in $\vec{\phi}$ we have
\begin{equation}
S^{(2)}_{eff}[\vec{\phi}] \approx\sum_{q}[\frac{U}{6}-\frac{2U^{2}}{9}\chi_{0}(q)]\vec{\phi}_{q}\cdot \vec{\phi}_{-q}+const
\end{equation}
Here $q=(\mathbf{q},i\omega_{n})$ is a more condensed notation for the wave vector and the Mastubara frequency of the fluctuating field $\vec{\phi}$, $const$ is an unimportant real constant. The bare susceptibility of the band electron is given by
\begin{equation}
\chi_{0}(q)=\mathrm{Tr}[G_{0} \mathbf{s}_{q}G_{0}\mathbf{s}_{-q}]
\end{equation} 
in which $\mathbf{s}_{q}$ denotes the Fourier component of the spin density operator at wave vector $\mathbf{q}$ and Mastubara frequency $i\omega_{n}$.
Using the property that $\chi_{0}(q)=\chi_{0}(-q)=(\chi_{0}(q))^{*}$ and that $\vec{\phi}_{-q}=\vec{\phi}^{*}_{q}$ we know that $S^{(2)}_{eff}[\vec{\phi}]$ is a real number and thus $e^{-S_{eff}[\vec{\phi}]}$ is positive-definite in the Gaussian limit.

More generally, using the fact that the matrix $G_{0}=[-(\partial_{\tau}-\mu)\delta_{i,j}+t_{i,j}]^{-1}$ is real and spin rotational invariant, it can be shown that all terms in the expansion in Eq.47 are real. However, this does not imply that $e^{-S_{eff}[\vec{\phi}]}$ is always positive-definite since the expansion in Eq.47 may not converge. In fact, the expansion of the matrix $\ln[\mathbf{I}+\frac{2U}{3}G_{0}\vec{\phi}_{i}\cdot \mathbf{s}_{i}]$ becomes ill-defined when $\mathbf{I}+\frac{2U}{3}G_{0}\vec{\phi}_{i}\cdot \mathbf{s}_{i}$ is not positive-definite.

An exact treatment of the negative sign problem is challenging. However, in line with the spirit of the random phase approximation, it is reasonable to argue that the field configuration with a non-positive-definite weight should be unimportant to the path integral as a result of the destructive interference of its fast fluctuating phase. An approximate action for the remaining field configurations with a positive-definite weight can be constructed in the same way as we derive the low energy effective theory of an interacting model. To the lowest order in $\vec{\phi}$, we can simply supplement the positive-definite Gaussian term with the usual local $\phi^{4}$ term, which is positive-definite. Thus, while it is in general impossible to solve exactly the negative sign problem in the Monte Carlo simulation of an interacting model, a simulation of an approximate action of its low energy physics is still possible. 

A more radical way to proceed is to start directly from a phenomenological guess of $S_{eff}[\vec{\phi}]$. A well known such example is the spin-fermion model proposed at the early stage of cuprate superconductivity study, in which $S_{eff}[\vec{\phi}]$ is assumed to take the following Gaussian form
\begin{equation}
S_{eff}[\vec{\phi}]=\frac{1}{2}\int_{0}^{\beta} d\tau\int_{0}^{\beta} d\tau'\sum_{\mathbf{q}} \chi^{-1}(\mathbf{q},\tau-\tau') \vec{\phi}_{\mathbf{q}}(\tau) \cdot\vec{\phi}_{-\mathbf{q}}(\tau')
\end{equation}
Here $\chi(\mathbf{q},\tau-\tau')$ denotes the generalized susceptibility of the local moment system. A widely adopted form for it is proposed by Millis, Monien and Pines(MMP) in the early 1990s\cite{Millis} and reads 
\begin{equation}
\chi(\mathbf{q},i\omega_{n})=\frac{\chi(\mathbf{Q})}{1+(\mathbf{q}-\mathbf{Q})^{2}\xi^{2}+|\omega_{n}|/\omega_{sf}}
\end{equation}
Here $\chi(\mathbf{Q})$ denotes the static susceptibility at the antiferromagnetic wave vector $\mathbf{Q}=(\pi,\pi)$. $\xi$ measures the correlation length of the antiferromagnetic fluctuation of the local moment. $\omega_{sf}$ is the characteristic frequency of the Landau damped fluctuation of the local moment in the background of itinerant quasiparticles. These parameters should all be understood as phenomenological variables to be determined from fitting experimental spin fluctuation spectrum. At a more microscopic level, the MMP form of $\chi$ can be derived from the expansion of the RPA susceptibility around an antiferromagnetic quantum critical point, in which $\chi(\mathbf{Q})\propto \xi^{2}$, $\omega_{sf}\propto\xi^{-2}$. 

An important property of the MMP susceptibility is that it is real in imaginary frequency. This actually holds more generally for any physical susceptibility, as can shown from its spectral representation. More specifically, we have
\begin{eqnarray}
\chi(\mathbf{q},i\omega_{n})&=&\frac{1}{2\pi}\int^{\infty}_{-\infty}d\omega' \frac{R(\mathbf{q},\omega')}{i\omega_{n}-\omega'}\nonumber\\
&=&-\frac{1}{\pi}\int_{0}^{\infty}\frac{\omega'R(\mathbf{q},\omega')}{\omega_{n}^{2}+\omega'^{2}}
\end{eqnarray}
Here we have used the property $R(\mathbf{q},\omega)=-R(\mathbf{q},-\omega)$ for the real spectral function. 
Thus a Monte Carlo simulation of the distribution $ e^{-S_{eff}[\vec{\phi}]}$ is free from the negative sign problem. In real computation, a $\phi^{4}$ term can be added to $S_{eff}[\vec{\phi}]$ to aid the stability of the simulation.

From such a simulation, we can obtain both $K$ and $C(i\omega)$ faithfully and thus the optical conductivity in imaginary frequency. To obtain the optical conductivity in real frequency, a Wick rotation $i\omega\rightarrow \omega+i0^{+}$ is needed. Such a Wick rotation can be done through numerical analytic continuation. Even though such a numerical procedure suffers from the ambiguity related to the exponential suppression in the Boltzmann weight at large imaginary time, $\sigma(i\omega_{n})$ along already contains important information that can be used as consistent check of the theoretical results.

\section{A case study: the optical conductivity in the presence of antiferromagnetic thermal spin fluctuation}  
The calculation simplifies greatly if we restricted ourself to the situation of thermal fluctuation of the local moment, as is considered in the study of Ref.[\onlinecite{Lin}]. In such a situation, there is no time dependence in the field $\vec{\phi}$ and the electron Green's function is diagonal in the Mastubara frequency. Denoting the single particle eigenstates of $\mathcal{H}_{\vec{\phi}}[\psi,\psi^{\dagger},\mathbf{A}=0]$ as $\psi^{m}_{i,\alpha}$ with the corresponding eigenvalue as $E_{m}$, we have
\begin{equation}
G_{i\alpha,j\alpha'}(i\nu_{n})=\sum_{m}\frac{\psi^{m}_{i\alpha}\ \psi^{m*}_{j\alpha'}}{i\nu_{n}-E_{m}}
\end{equation}
in which $i\nu_{n}=(2n+1)\pi k_{B}T$ is the fermionic Mastubara frequency. We thus have
\begin{eqnarray}
K&=&T\langle \mathrm{Tr}[G\mathbf{K}]\rangle\nonumber\\
&=&\langle T\sum_{i\nu_{n}}\sum_{i,j,\alpha}G_{i\alpha,j\alpha}(i\nu_{n})\mathbf{K}_{j,i}  \rangle\nonumber\\
&=&\langle \ \sum_{i,j,m,\alpha}f(E_{m})\ \psi^{m*}_{j\alpha}\mathbf{K}_{j,i}\psi^{m}_{i\alpha}\ \rangle
\end{eqnarray}
The paramagnetic kernel $C(i\omega_{n})$ can be found similarly. Noting that 
\begin{equation}
\mathrm{Tr}[G\mathbf{j}(i\omega_{n})]=0
\end{equation}
for $i\omega_{n}\neq 0$ since $\vec{\phi}$ is now time independent, we have
\begin{eqnarray}
C(i\omega_{n})=T\langle \mathrm{Tr}[G\mathbf{j}(i\omega_{n})G\mathbf{j}(-i\omega_{n})]\rangle
\end{eqnarray}
in which
\begin{eqnarray}
&&T \mathrm{Tr}[G\mathbf{j}(i\omega_{n})G\mathbf{j}(-i\omega_{n})]\nonumber\\
&=&T\sum_{i\nu_{n}}\sum_{i,j,k,l,\alpha\alpha'}G_{i\alpha,j\alpha'}(i\nu_{n}+i\omega_{n})\mathbf{j}_{j,k}G_{k\alpha',l\alpha}(i\nu_{n})\mathbf{j}_{l,i}\nonumber\\
&=&T\sum_{i\nu_{n}}\sum_{m,m'}\frac{\mathbf{j}^{2}_{m,m'}}{(i\nu_{n}+i\omega_{n}-E_{m})(i\nu_{n}-E_{m'})}\nonumber\\
&=&\sum_{m,m'}\mathbf{j}^{2}_{m,m'}\frac{f(E_{m'})-f(E_{m})}{i\omega_{n}-(E_{m}-E_{m'})}
\end{eqnarray}
Here 
\begin{equation}
\mathbf{j}^{2}_{m,m'}=\sum_{i,j,k,l,\alpha\alpha'}(\psi^{m*}_{j\alpha'}\mathbf{j}_{j,k}\psi^{m'}_{k\alpha'})\times(\psi^{m'*}_{l\alpha}\mathbf{j}_{l,i}\psi^{m}_{i\alpha})
\end{equation}

After the analytic continuation, we find that the regular part of the optical conductivity is given by
\begin{eqnarray}
\sigma^{reg}(\omega)&=&-\frac{\mathrm{Im} C^{R}(\omega+i0^{+})}{\omega}\nonumber\\
&=&\langle \  \frac{\pi}{\omega}\sum_{m,m'}\mathbf{j}^{2}_{m,m'}\times[f(E_{m'})-f(E_{m})]\nonumber\\
 &\times&\delta(\omega-[E_{m}-E_{m'}])  \ \rangle
\end{eqnarray}
We note that $\sigma^{reg}(\omega)$ should be understood as a rank-2 tensor. Using the rotational symmetry of the system it can be shown that $\sigma^{reg}(\omega)$ is diagonal and $\sigma^{reg}_{xx}(\omega)=\sigma^{reg}_{yy}(\omega)$. Here we will focus on $\sigma^{reg}_{xx}(\omega)$.

In the following, we apply this formula to study the effect of thermal spin fluctuation on the optical conductivity of the cuprate superconductors. We will use the same form of the spin susceptibility as adopted in Ref.[\onlinecite{Lin}], which reads
\begin{equation}
\chi(\mathbf{q},i\omega_{n})=\frac{\chi_{0}}{\xi^{-2}+(\mathbf{q}-\mathbf{Q})^{2}}\delta_{i\omega_{n},0}
\end{equation}
This amounts to take the limit $\frac{\omega_{sf}}{k_{B}T}\rightarrow 0$ in Eq. 51. The effective action $S_{eff}[\vec{\phi}]$ then takes the form of
\begin{equation}
S_{eff}[\vec{\phi}]=\frac{\beta}{2}\sum_{i,j}\chi^{-1}_{i,j}\vec{\phi}_{i}\cdot\vec{\phi}_{j}
\end{equation}
with
\begin{equation}
\chi^{-1}_{i,j}=\frac{1}{N}\sum_{\mathbf{q}}\chi ^{-1}(\mathbf{q})e^{i\mathbf{q}\cdot(\mathbf{r}_{i}-\mathbf{r}_{j})}
\end{equation}
To perform numerical simulation we now formulate the problem on a finite square lattice with $L\times L$ lattice sites. Periodic boundary condition will be assumed in both dimensions. To be compatible with such a lattice regulation, we make the following replacement in the phenomenological susceptibility Eq.60
\begin{equation}
(\mathbf{q}-\mathbf{Q})^{2}\rightarrow 4+2[\cos(q_{x})+\cos(q_{y})]
\end{equation}
We then have
\begin{equation}
\chi^{-1}_{i,j}=\frac{1}{\chi_{0}}[(4+\xi^{-2})\delta_{i,j}+\delta_{\mathbf{r}_{i}-\mathbf{r}_{j},\bm{\delta}}]
\end{equation}
in which $\bm{\delta}$ denotes the four nearest-neighboring vectors on the square lattice. The effective action for $\vec{\phi}$ then becomes
\begin{equation}
S_{eff}[\vec{\phi}]=\frac{\beta(4+\xi^{-2})}{2\chi_{0}}\sum_{i}\vec{\phi}_{i}\cdot\vec{\phi}_{i}+\frac{\beta}{2\chi_{0}}\sum_{i,\bm{\delta}}\vec{\phi}_{i}\cdot\vec{\phi}_{i+\bm{\delta}}
\end{equation}

\begin{figure}
\includegraphics[width=7cm]{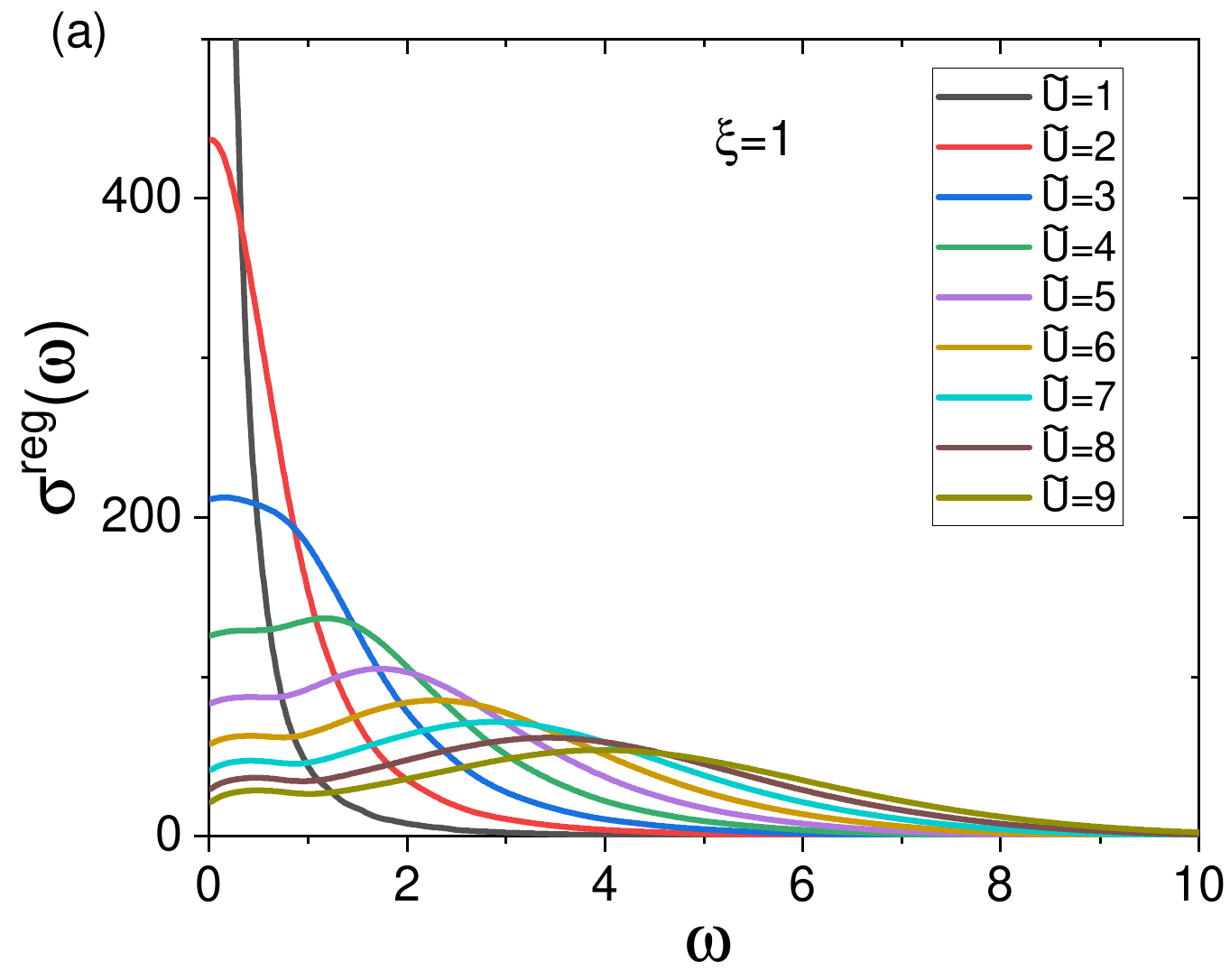}
\includegraphics[width=7cm]{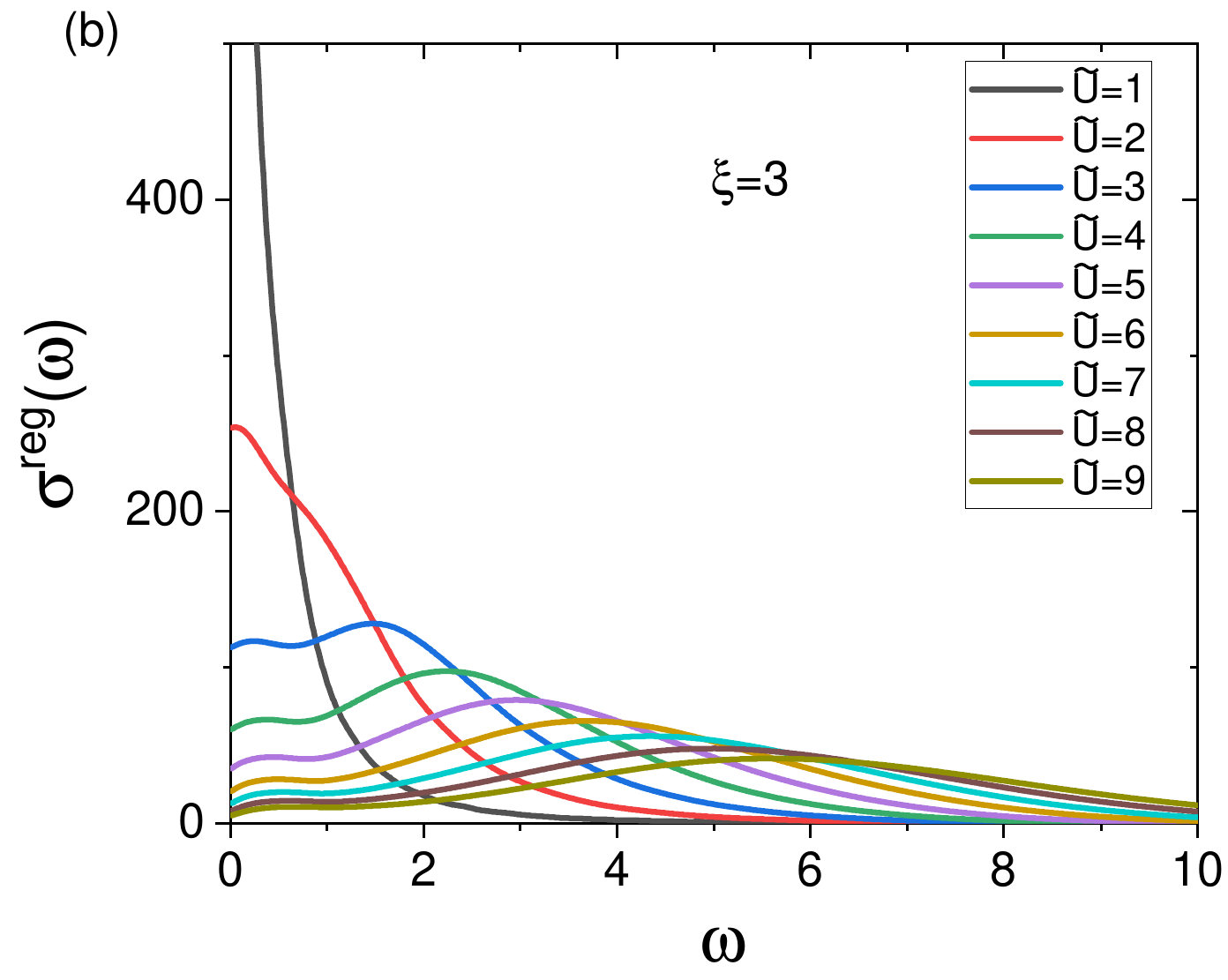}
\includegraphics[width=7cm]{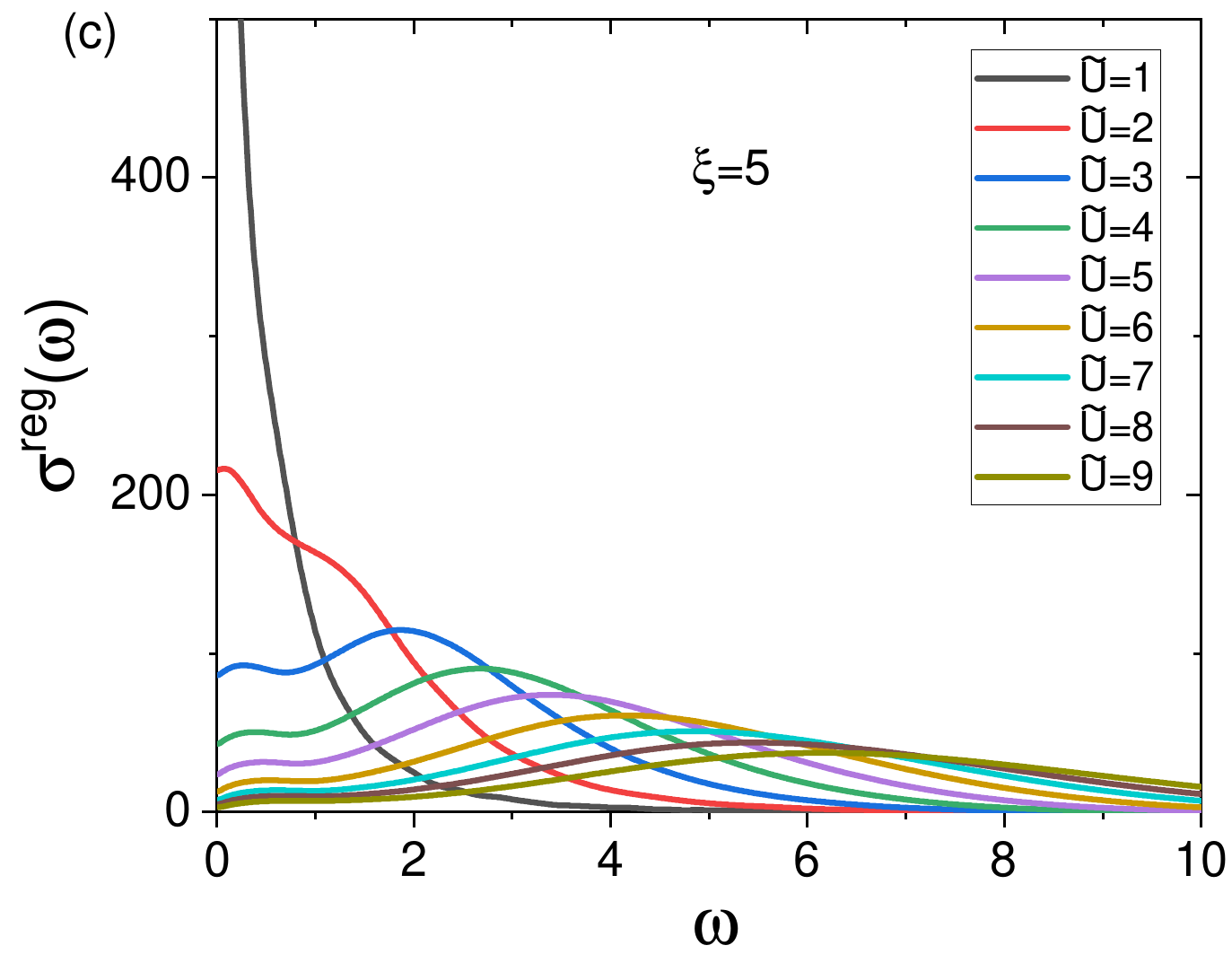}
\includegraphics[width=7cm]{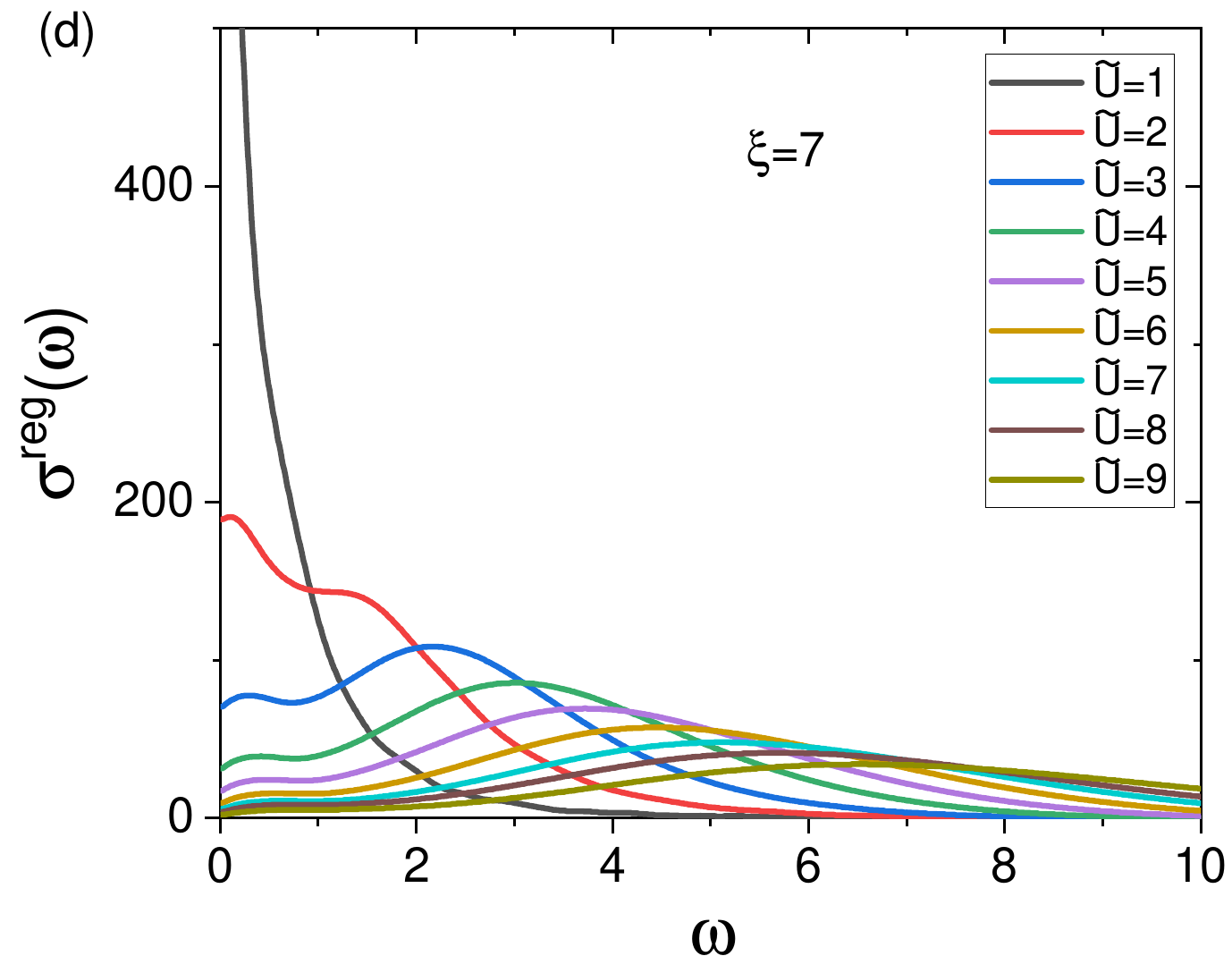}
\caption{The optical conductivity of the system when we assume an action of the form Eq.65 for the thermally fluctuating moment. Here we have treated $\tilde{U}$ and $\xi$ as two independent parameters, although in reality both $\tilde{U}$ and $\xi$ are temperature dependent quantities. The energy is measured in unit of $t$ and and we have set $k_{B}T=0.1t$ for the electron. The $\delta$-function peak is broadened into a Lorentzian peak of width $0.03t$ in the calculation. The calculation is done on a lattice with $L\times L=400$ sites.}
\end{figure}  

We perform standard Monte Carlo sampling over the distribution $e^{-S_{eff}[\vec{\phi}]}$ with the Metropolis algorithm. Since the three components of $\vec{\phi}$ fluctuate independently in the Gaussian limit, the sampling reduces to that of three independent Gaussian distributions. At the same time, since the parameter $\beta\chi^{-1}_{0}$ in $S_{eff}[\vec{\phi}]$ can be absorbed into a redefinition of the integration variable $\vec{\phi}$, with the only effect being replacing $U$ by $\tilde{U}=U\sqrt{\chi_{0}/\beta}$ in $\mathcal{H}_{\vec{\phi}}[\psi,\psi^{\dagger},\mathbf{A}=0]$, we will set $\beta\chi^{-1}_{0}=1$ and treat $\tilde{U}$ as a free parameter in the following. $\tilde{U}$ measures the overall strength of the coupling between the antiferromagnetic spin fluctuation and the electron. Another free parameter in our calculation is the spin correlation length $\xi$. We will consider a tight binding model with only the nearest and the next-nearest neighboring hopping integrals. The band parameters are set to be $t'=-0.3t$, with $t$ and $t'$ the hopping integrals between the nearest and the next-nearest-neighboring sites. In the following, we will use $t$ as the unit of energy.

We have generated 6400 statistically independent thermal fluctuation configurations $\vec{\phi}$ from the distribution $e^{-S_{eff}[\vec{\phi}]}$ using the Metropolis algorithm. The optical conductivity of the system is computed by averaging the optical conductivity for each of these 6400 configurations. Here we will be satisfied with a qualitative understanding of the physical consequence of the thermal spin fluctuation, rather a quantitative comparison with the detailed doping and temperature dependence of the measured optical spectrum in the cuprate superconductors. We thus perform the calculation at a typical temperature of $k_{B}T=0.1t$ and set chemical potential to be $\mu=-t$, which corresponds to a doping level of $15\%$. All calculations are performed on a lattice with $L\times L=400$ sites and with periodic boundary condition in both the $x$ and the $y$-direction.

The calculated optical conductivity is shown in Fig.1 for various set of parameters. In our calculation, we have treated $\tilde{U}$ and $\xi$ as two independent parameters, although in reality both of them are temperature dependent quantities. The results presented in Fig.1 should thus only be understood at a qualitative ground. For small $\tilde{U}$, $\sigma^{reg}(\omega)$ is seen to be dominated by a Drude peak at low energy. This is to be contrasted with the behavior at large $\tilde{U}$, in which case the optical spectrum is dominated by a broad mid-infrared peak extending to the energy scale of the band width. In between these two limits, the optical spectrum exhibits a two-component structure with a Drude peak at low frequency and a mid-infrared peak at higher frequency. Such a two-component character becomes increasingly more evident with the increase of the correlation length of the local moment fluctuation. We find that the low energy Drude peak can be attributed to the residual electron density of state near the fermi level in the presence of the thermal spin fluctuation.   

A two-component structure has been indeed observed in the optical spectrum of the cuprate superconductors\cite{Basov,Marel,Heumen}. However, the Drude peak predicted here is much too weak to be consistent with the experimental observations. Such a discrepancy should be attributed to the neglect of the quantum nature of the local moment fluctuation. In fact, the neglect of $\tau$-dependence in $\vec{\phi}_{i}(\tau)$ becomes invalid for electron transition below some characteristic energy of the order of $\omega_{sf}$, whence the quantum nature of the spin fluctuation becomes important. This is particularly the case in the hole-doped cuprate superconductors since its spin fluctuation is more dynamical than that in the electron-doped cuprate superconductors. Such a quantum effect is expected to recover partially the electron density of state near the fermi level from the SDW gapping caused by the thermal spin fluctuation and is thus expected to enhance the spectral weight contained in the Drude peak. On the other hand, the mid-infrared weight is not expected to be significantly influenced by such an effect since it is located at an energy significantly higher than $\omega_{sf}$. From the discussion presented in the last section, we know that an exact treatment of such a quantum effect is possible with sign-problem-free quantum Monte Carlo simulation in the spin-fermion model framework. Such a study will be pursued in an independent work.

To be complete, we also present the result for the optical conductivity when we assume a non-Gaussian action off the following form
\begin{equation}
S_{eff}[\vec{\phi}]=\eta \sum_{i,\bm{\delta}}\vec{\phi}_{i}\cdot\vec{\phi}_{i+\bm{\delta}}
\end{equation}
The non-Gaussian nature of this action is encoded in the following constraint on the fluctuating field $\vec{\phi}_{i}$ 
\begin{equation}
\vec{\phi}_{i}\cdot\vec{\phi}_{i}=1
\end{equation}
$\eta$ is a constant introduced to tune the correlation length of the fluctuating moment. Such an action describes the thermal fluctuation in the renormalized classical regime of a Heisenberg antiferromagnet\cite{CHN}. This is thought to be relevant to the situation of the electron-doped cuprate superconductors\cite{Damascelli}, in which the spin fluctuation is more long-range correlated and more static. We have sampled such a distribution with the heat bath algorithm supplemented by the over-relaxation update on the unit vector $\vec{\phi}_{i}$\cite{Young,Okubo}. The desired correlation length $\xi$ is achieved by tuning the value of the parameter $\eta$. The obtained optical conductivity is shown in Fig.2 for various value of $\tilde{U}$ and $\xi$. When compared to the result we got for the Gaussian action above, the two-component character in the optical spectrum becomes even more evident. It is interesting to compare these results with the observations in the electron-doped cuprate superconductors.

\begin{figure}
\includegraphics[width=7cm]{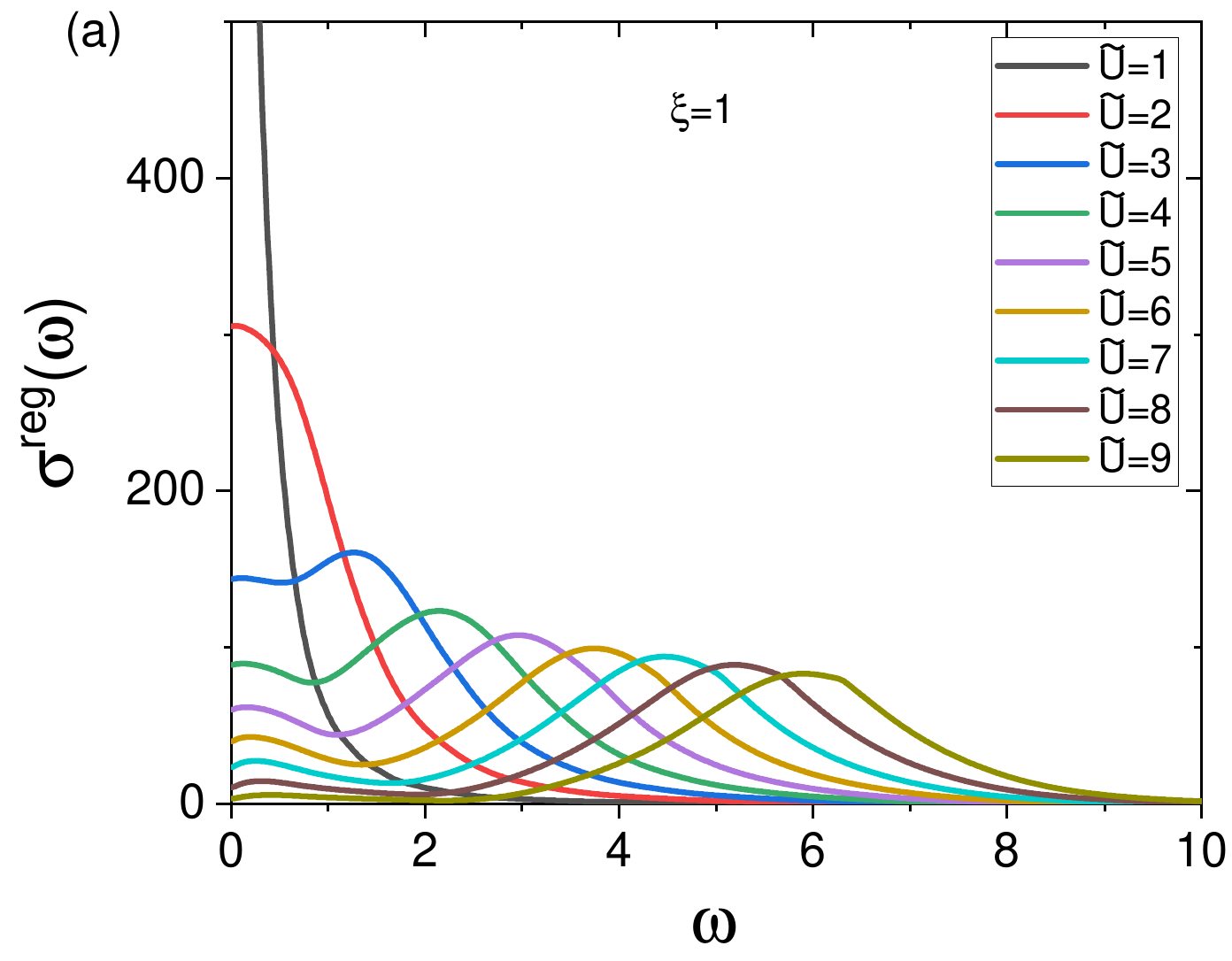}
\includegraphics[width=7cm]{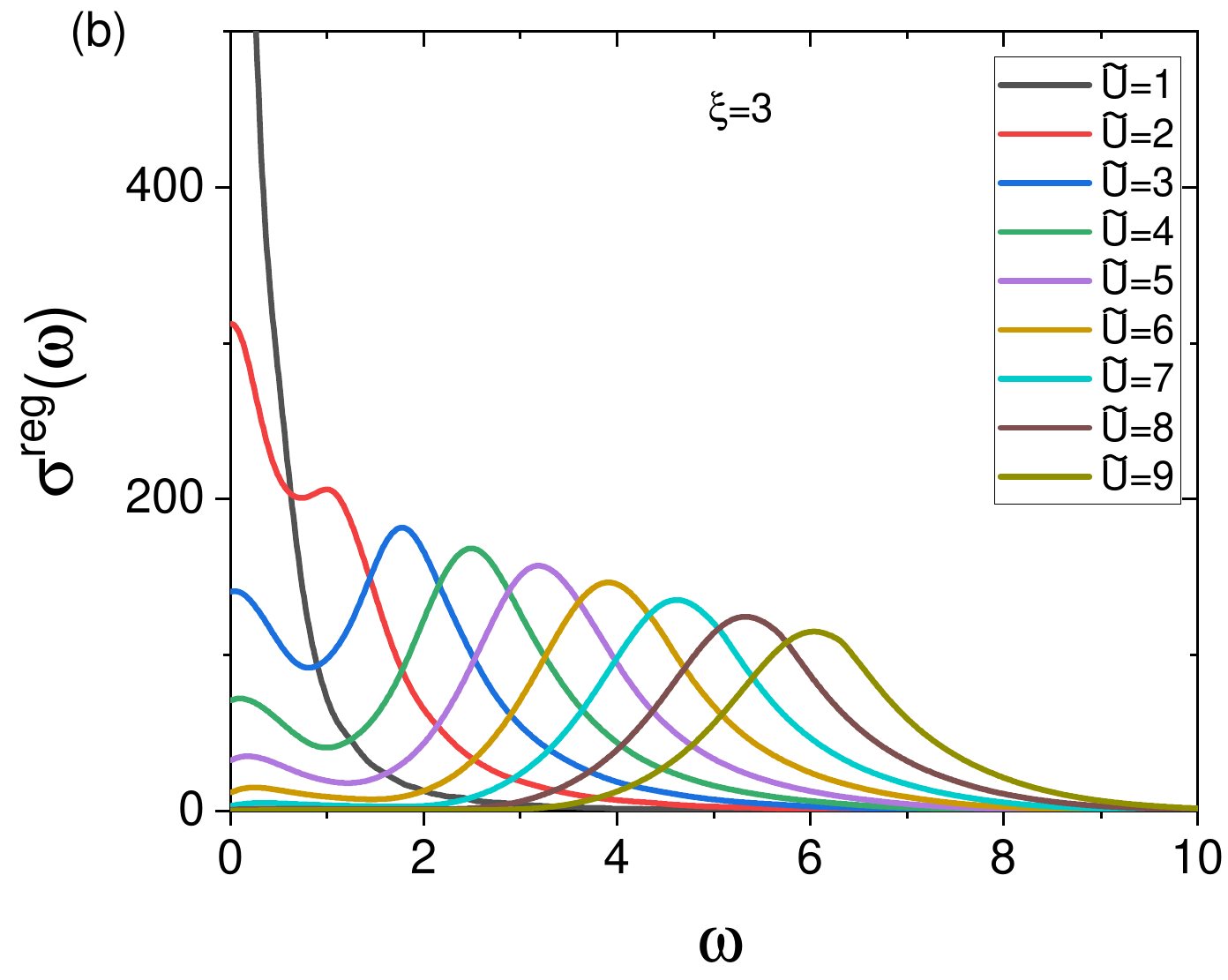}
\includegraphics[width=7cm]{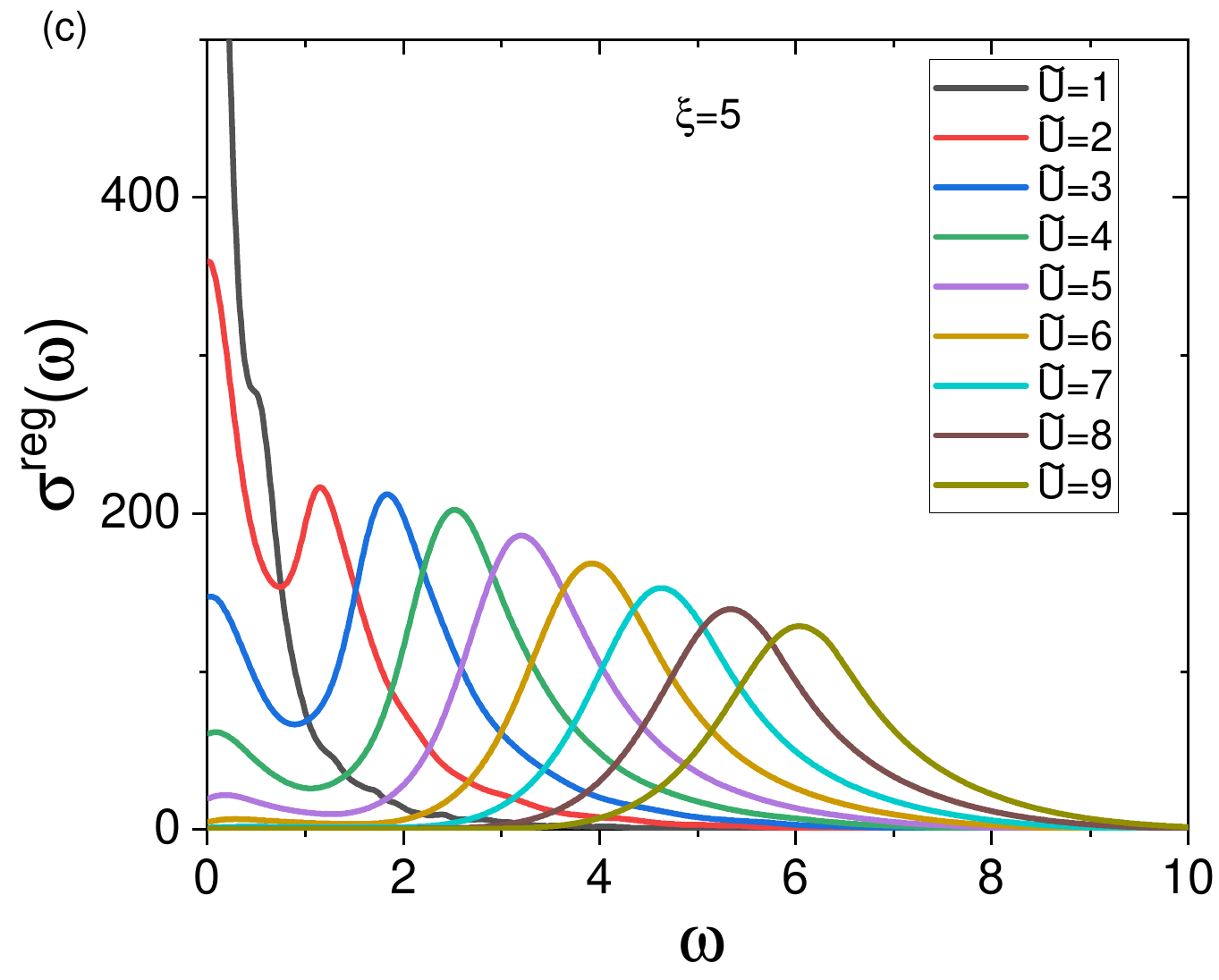}
\includegraphics[width=7cm]{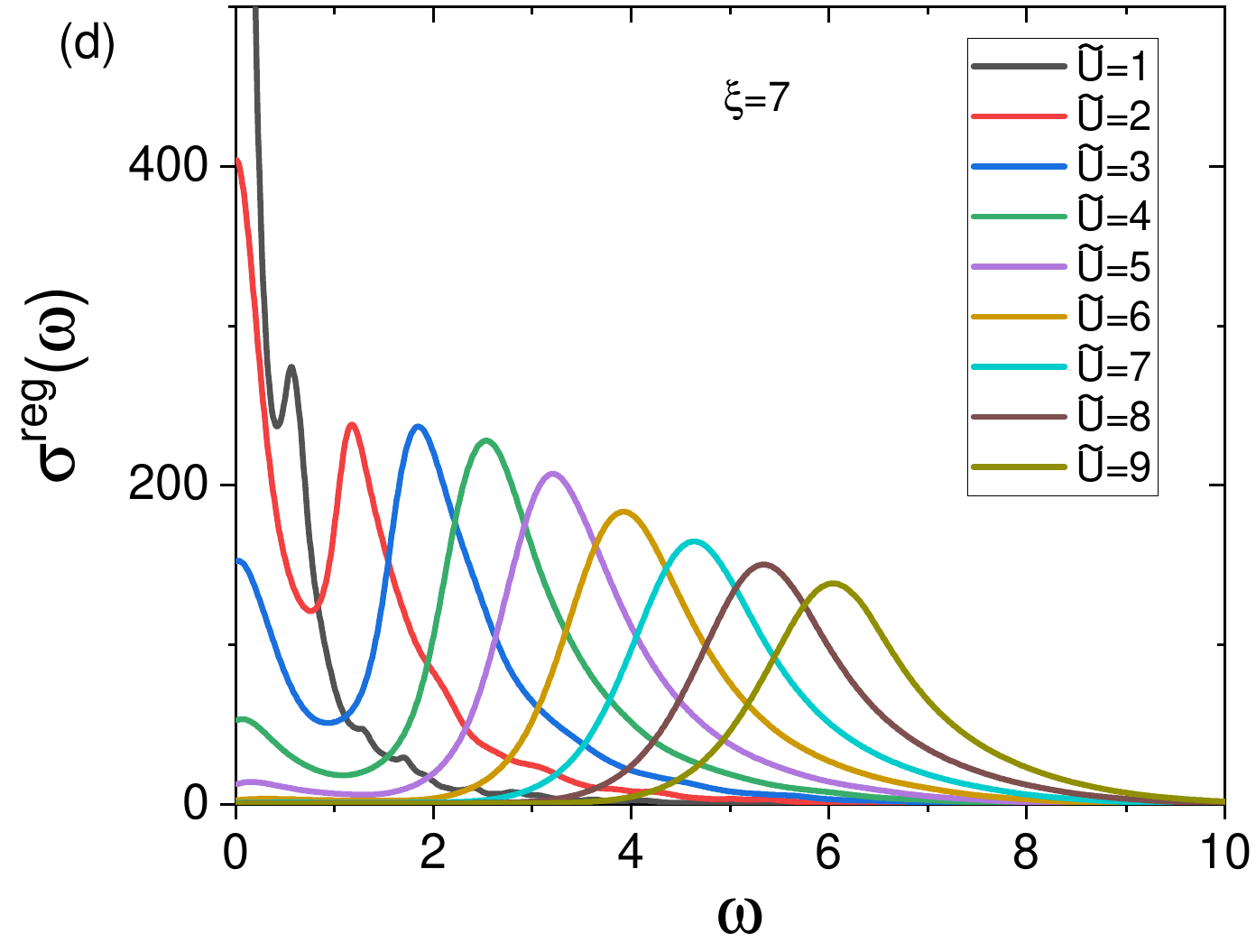}
\caption{The optical conductivity of the system when we assume a non-Gaussian action of the form Eq.66 for the thermal spin fluctuation. Such an action describes the thermal fluctuation in the renormalized classical regime of a Heisenberg antiferromagnet. It is thought to be relevant to the electron-doped cuprate superconductors. The $\delta$-function peak is broadened into a Lorentzian peak of width $0.03t$. The calculation is done on a lattice with $L\times L=400$ sites.}
\end{figure}  

\section{Conclusions and discussions}
In this paper, we have derived an exact formula for the optical conductivity of the Hubbard model. We find that the optical conductivity of the system can be represented as the ensemble average of the optical conductivity of non-interacting systems coupled to fluctuating moments. We find that the notorious negative sign problem in the Monte Carlo sampling of the distribution of such fluctuating moments disappears in two important limits. The first is the high temperature limit in which only thermal spin fluctuation matters. The second is the Gaussian limit which becomes valid when we are concerned with the weakly correlated regime of the system. 

It is interesting to note that the occurrence of the negative sign problem is directly related to the nonlinearity effect in the spin fluctuation, which again is closely related to the formation of local moment in the system. This implies the possibility of solving the negative sign problem by finding a sign-problem-free approximation for the non-Gaussian action of the fluctuating local moment that preserve the essential physics of the problem. We find that the conventional $\phi^{4}$ approximation of the non-Gaussian term just fulfills such a need. We thus think that the combination of the low energy effective theory consideration and the standard quantum Monte Carlo simulation can be very fruitful in the numerical study of strongly correlated electron models. A prominent example of this kind is the spin-fermion model, for which we show that the quantum Monte Carlo simulation is free from the negative sign problem. 

In this paper, we have applied our formula to study the effect of the thermal spin fluctuation on the optical conductivity of the cuprate superconductors. Using a Millis-Monien-Pines type Gaussian effective action for the fluctuating moment, we have simulated how the optical conductivity evolves with the correlation length $\xi$ and the overall strength of the local moment fluctuation $\tilde{U}$. The optical conductivity calculated from our numerical simulation is found to exhibit a two-component structure, with a Drude component at low frequency and a mid-infrared component at higher frequency. Such a two-component character is found to become more evident as we increase the correlation length. While such a two-component phenomenology is also observed in the optical spectrum of the cuparte superconductors, the Drude weight predicted here is too weak to be consistent with the experimental observations. We think that such a discrepancy should be attributed to the neglect of quantum nature of the local moment fluctuation at low frequency, which acts to recover partially the electron density of state around the fermi level from the SDW gapping caused by the thermal spin fluctuation. Such an effect is expected to be more important in the hole-doped cuprate superconductors than in the electron-doped cuprate superconductors, since the local moment fluctuation is more dynamical in the former($\omega_{sf}$ is larger). We note that an exact treatment of such quantum effect is possible with sign-problem-free quantum Monte Carlo simulation within the spin-fermion model framework. A study of such an effect will be pursued in a future work.

In addition, we find that the two-component character in the optical spectrum becomes more evident when we assume a non-Gaussian action for the fluctuating moment that describe the thermal spin fluctuation in the renormalized classical regime of a Heisenberg antiferromagnet. Such an action is thought to be relevant to the electron-doped cuprate superconductors, in which the antiferromagnetic spin fluctuation is more long-ranged correlated and more static. A semi-quantitative comparison of the result presented in Fig.2 with the optical spectrum of the electron-doped cuprate superconductors is thus very likely to be realistic.

The computational scheme presented in this paper is certainly not restricted to the study of the optical conductivity of the cuprate superconductors. A more interesting quantity to be calculated is the Hall response of the system. From the early days of the high-T$_{c}$ era it has been well known that Hall number $1/R_{H}$ of the cuprate superconductors exhibits anomalous temperature dependence in the so called strange metal regime of the phase diagram. More recently, it is found that $1/R_{H}$ undergoes sharp transition from the $1/R_{H}\propto x$ to $1/R_{H}\propto 1+x$ behavior around the so called pseudogap end point $x^{*}$, where quantum critical behavior of unknown origin has been found in both transport and thermodynamical measurements\cite{Taillefer,Proust,Hussey}. While such a transition in $1/R_{H}$ seems to suggest a change in the fermi surface topology of the system, no evidence of symmetry breaking phase transition is found at $x^{*}$. As a possible solution to this puzzle, it has been conceived that short range antiferromagnetic fluctuation may have a similar effect as the long range antiferromagnetic order on the transport behavior of the system. It is thus interesting to see if the simulation of spin-fermion model can reproduce the qualitative features of the experimental observations. We note that the computation of the Hall response is significantly more complicated than the computation of the optical conductivity, since it involves the expansion of the effective action to higher order in the electromagnetic potential. The detailed result of such an attempt will be reported in a separate paper in the near future.  

\begin{acknowledgments}
We acknowledge the support from the National Natural Science Foundation of China(Grant No.12274457). 
\end{acknowledgments}

\end{document}